\documentstyle[graphicx,aps,amstex]{revtex}

\begin{document}
\draft

\draft
\preprint{ }
\title{Kinetic Theory of Collective Modes in Atomic Clouds above
the Bose-Einstein Transition Temperature}
\author{U. Al Khawaja$^{1}$,  C. J. Pethick$^{2}$ and H. Smith$^{1}$}
\address{$^1$\O rsted Laboratory, H. C. \O rsted Institute,
Universitetsparken 5, DK-2100 Copenhagen \O, Denmark\\
$^2$Nordita, Blegdamsvej 17, DK-2100 Copenhagen \O, Denmark.}
\date{\today}
\maketitle
\begin{abstract}
We calculate frequencies and damping rates of the lowest collective modes of
a dilute Bose gas confined in an anisotropic trapping potential above the
Bose-Einstein transition temperature. From the Boltzmann equation with a
simplified collision integral we derive a general dispersion relation that
interpolates between the collisionless and hydrodynamic regimes. In the case
of axially symmetric traps we obtain explicit expressions for the
frequencies and damping rates of the lowest modes in terms of a
phenomenological collision time. Our results are compared with microscopic
calculations and experiments.
\end{abstract}

\pacs{PACS numbers: 03.75.Fi,03.65.Db,05.30.Jp,32.80.Pj}


\section{Introduction}

Above the Bose-Einstein transition temperature both collisionless and
hydrodynamic modes can exist in a trapped gas. In the collisionless regime
the frequency of the mode is large compared with the interatomic collision
frequency, and the mean free path is larger than the wavelength of the mode
and the dimension of the cloud. In the hydrodynamic regime the opposite is
true. Previous theoretical studies have focused on the collisionless
\cite{{george2},{stoof}} and hydrodynamic regimes
\cite{{notes},{george1},{5string},{6string},{griffin2}}. Analysis of recent
measurements of
frequencies and damping rates of collective modes in trapped Bose gases
above the Bose-Einstein transition temperature
\cite{{ket},{ket1},{jila1},{jila2}} indicates that the experiments were
performed
under conditions intermediate between the collisionless and hydrodynamic
regimes \cite{{george2},{notes}}, and in 
Ref. \cite{george2} a simple interpolation formula for the intermediate
regime was proposed. The aim of this paper is to calculate frequencies and
damping rates of modes in the intermediate regime, and to examine the
validity of the interpolation formula.

The difficulty in this problem arises from the inhomogeneity of the system.
Unlike the case of homogeneous systems conditions may be hydrodynamic in the
center of the cloud and collisionless near its surface. In a trapped gas the
collisionless limit is always well-defined in the sense that the frequency
may exceed the collision rate everywhere, but hydrodynamic conditions can
only be achieved in a limited region of space, not in the outer parts of the
cloud. Therefore one would expect a single relaxation-time approximation to
work less well than for a homogeneous system.

At temperatures above the transition temperature the gas is dilute in the
sense that the mean interatomic spacing is much larger than the interatomic
scattering length. Mean-field effects can be neglected since the potential
energy due to interactions, $nU_{0}$, is much less than the thermal energy $%
k_{B}T$. Here $n$ is the number density, $U_{0}$ is the effective two-body
interaction (which is given in terms of the scattering length $a$ as $%
U_{0}=4\pi a\hbar ^{2}/m,$ where $m$ is the atomic mass), $k_{B}$ is the
Boltzmann constant, and $T$ is the temperature. Then the
atoms can be described by a distribution function that obeys the Boltzmann
equation \cite{george2}. We solve the Boltzmann equation with a trial
solution appropriate to the collective modes being studied, using a
relaxation time approximation to the collision integral.

In the next section we derive a general relation for mode frequencies
that interpolates between the collisionless and hydrodynamic regimes in an
anisotropic trap. In section \ref{axial} we consider the experimentally
relevant case of axially symmetric traps and calculate frequencies and
damping rates of the lowest modes. In section \ref{disc} we discuss the
validity of our approach by comparing our results with microscopic
calculations and with experiment. Section \ref{conc} contains a summary of
our conclusions.

\section{The Boltzmann Equation and Collective Modes}

\label{disp} In this section we derive the dispersion relation of the lowest
modes of a Bose gas confined by an anisotropic harmonic potential given by 
\begin{equation}
V({\mathbf{r}})=\frac{1}{2}m(\omega _{1}^{2}x^{2}+\omega _{2}^{2}y^{2}+\omega
_{3}^{2}z^{2}),  \label{1}
\end{equation}
where $\omega _{1},$ $\omega _{2},$ and $\omega _{3}$ are the characteristic
frequencies of the trap in the three directions.

The dynamics of the dilute Bose gas above the transition temperature can be
described by a semi-classical distribution function $f({\mathbf{p,r}},t),$
where $\mathbf{p}$ is the particle momentum. The distribution function
satisfies the Boltzmann equation 
\begin{equation}
\frac{\partial f}{\partial t}+\frac{1}{m}{\mathbf{p\cdot \nabla }}_{\mathbf{r}%
}f-{\mathbf{\nabla _{r}}}V\cdot {\mathbf{\nabla }}_{\mathbf{p}}f=\left( \frac{%
\partial f}{\partial t}\right) _{coll.}.  \label{2}
\end{equation}
Here $\left( \partial f/\partial t\right) _{coll.}$\ is the contribution of
collisions to the rate of change of $f$. For two-body scattering it is given
by 
\begin{eqnarray}
\left( \frac{\partial f({\mathbf{p,r},t)}}{\partial t}\right) _{coll.}\  &=&-{%
\frac{1}{(2\pi \hbar )^{3}}}\int d{\mathbf{p}}_{1}\int d\sigma \left|
{\mathbf{v%
}}-{\mathbf{v}}_{1}\right| \left[ f({\mathbf{p,r}})f({\mathbf{p}}_{1}
{\mathbf{,r}}%
)(1+f({\mathbf{p}}{\acute{}}{\mathbf{,r}}\;))(1+f({\mathbf{p}}{\acute{}}_{1}%
\mathbf{,r}))\right.   \nonumber \\
&&\left. -f({\mathbf{p}}{\acute{}}{\mathbf{,r}})\ f({\mathbf{p}}{\acute{}}%
_{1}{\mathbf{,r}})(1+f({\mathbf{p,r}}))(1+f({\mathbf{p}}_{1}{\mathbf{,r}}))
\right] .
\label{3}
\end{eqnarray}
The collision integral involves scattering processes with two incoming
particles of momenta $\mathbf{p}$ and ${\mathbf{p}}_{1}$ (and velocities $%
\mathbf{v}$ and ${\mathbf{v}}_{1}$) and two outgoing particles of momenta $%
{\mathbf{p}}{\acute{}}$ and ${\mathbf{p}}{\acute{}}_{1},$ respectively. The
differential cross section is denoted by $d\sigma$.

We shall linearize the Boltzmann equation in small deviations of the
distribution function from the equilibrium one. This can be achieved simply
by inserting the following expression 
\begin{equation}
f({\mathbf{p,r}})=f^{(0)}({\mathbf{p,r}})+f^{(0)}({\mathbf{p,r}})(1+f^{(0)}(%
{\mathbf{p,r}}))\psi ({\mathbf{p,r}})  \label{exp1}
\end{equation}
in Eqs. (\ref{2}) and (\ref{3}), where $\psi ({\mathbf{p,r}})$ describes the
deviation from equilibrium, and $f^{(0)}$ is the equilibrium
distribution function 
\begin{equation}
f^{(0)}({\mathbf{p,r}})
=\left\{ \exp\left[(p^{2}/2m+V({\mathbf{r}})-\mu )
/k_{B}T\right]-1\right\} ^{-1}.
\end{equation}
To first order in $\psi ,$ the Boltzmann equation
assumes the following form 
\begin{equation}
\ \left( \frac{\partial }{\partial t}+\frac{1}{m}{\mathbf{p\cdot \nabla }}_{%
\mathbf{r}}-{\mathbf{\nabla _{r}}}V\cdot {\mathbf{\nabla }}_{\mathbf{p}}\right)
\psi ({\mathbf{p,r}})=-I[\psi ({\mathbf{p,r}})],  \label{8}
\end{equation}
where $I$ is a linear operator given by 
\begin{eqnarray}
I[\psi ] &=&{\frac{1}{(2\pi \hbar )^{3}}}\int d{\mathbf{p}}_{1}\int d\sigma
\left| {\mathbf{v}}-{\mathbf{v}}_{1}\right| \ f^{(0)}({\mathbf{p}}_{1}
{\mathbf{,r}}%
)(1+f^{(0)}({\mathbf{p}}{\acute{}}{\mathbf{,r}}))(1+f^{(0)}({\mathbf{p}}
{\acute{}%
}_{1}{\mathbf{,r}})){\frac{1}{1+f^{(0)}
({\mathbf{p,r}})}}\times  \\
&&\left[ \psi ({\mathbf{p,r}})+\psi
({\mathbf{p}}_{1}{\mathbf{,r}})-\psi ({\mathbf{p%
}}{\acute{}}
{\mathbf{,r}})-\psi ({\mathbf{p}}{\acute{}}_{1}{\mathbf{,r}})\right] .
\end{eqnarray}

Since the scattering processes considered here conserve the number of
particles, momentum, and energy \cite{henrik book}, the zeroth, first, and
second moments of the collision integral, calculated by multiplying it by $1,
$ $\mathbf{p,}$ and $p^{2}$, respectively, and then integrating over $%
\mathbf{p,}$ must vanish. The quantities $1,$ $\mathbf{p,}$ and $p^{2}$ are
called the collision invariants.

\subsection{Dispersion Relation for Low-Lying Modes}

\label{sub1} The lowest collective mode corresponds to the center of mass
motion (the dipole mode). The frequency of this mode equals $\omega _{i}$ $%
(i=1,2,3)$ and does not depend on interatomic collisions. Therefore this
mode will not be considered further in this paper. We shall instead focus on
the higher modes which have a frequency $2\omega _{i}$ in the collisionless
limit. In the hydrodynamic limit one can show that the fluid velocity $%
\mathbf{u}$ associated with these modes is given generally by ${\mathbf{u}}%
=(ax,by,cz)$, where $a$, $b$, and $c$ are constants
\cite{{notes},{george1},{5string},{6string},{griffin2}}.
This expression is general in the sense that it does not depend on
temperature or the anisotropy of the trap. In spherical traps these modes
correspond to the monopole mode (or breathing mode) characterized by $a=b=c$%
, and the quadrupole modes for which $a+b+c=0$.

Let us attempt to find a solution $\psi $ that corresponds to the lowest
modes of the system described above. The deviation function $\psi $
corresponding to a flow with drift velocity $\mathbf{u}$ is proportional to $%
\mathbf{u}\cdot \mathbf{p}$. Acting on $\mathbf{u}\cdot \mathbf{p}$ with the
left hand side of Eq. (\ref{8}) will thus introduce terms like $x^{2}$, $%
p_{x}^{2} $ etc., and therefore we adopt as our trial function the general
form 
\begin{equation}
\psi =e^{-i\omega
t}[a_{1}x^{2}+b_{1}xp_{x}+c_{1}p_{x}^{2}+a_{2}y^{2}+b_{2}yp_{y}+c_{2}p_{y}^{2}
+a_{3}z^{2}+b_{3}zp_{z}+c_{3}p_{z}^{2}].
\label{5}
\end{equation}
Here the terms $x^{2},$ $y^{2},$ and $z^{2}$ correspond to changes in the
number of particles, and $xp_{x},$ $yp_{y},$ and $zp_{z}$ correspond to
changes in the local momentum density, and hence all these terms are
collision invariants. On the other hand the terms $p_{x}^{2},$ $p_{y}^{2},$
and $p_{z}^{2}$ are not collision invariants since only their sum $%
p^{2}=p_{x}^{2}+p_{y}^{2}+p_{z}^{2},$ which is proportional to the kinetic
energy, is. Mode damping therefore arises from only these latter terms.

In the relaxation time approximation the collision operator is associated
with a mean relaxation time $\tau (\mathbf{r)}$, which depends on the
position $\mathbf{r}$. One is restricted, however, in that for any
approximate expression for the collision operator $I,$ the above-mentioned
conservation laws should be satisfied. This can be ensured by requiring the
collision invariants to be eigenfunctions of the collision operator with
eigenvalue zero. However, for terms such as $p_{x}^{2}$, which are not
collision invariants and in momentum space have the symmetry of $Y_{2}^{m}$,
where $Y_{l}^{m}$ are the spherical harmonics, we have 
\begin{equation}
I[p_{x}^{2}]=-\frac{1}{\tau (\mathbf{r)}}(p_{x}^{2}-p^{2}/3),  \label{7}
\end{equation}
and similarly for $p_{y}^{2}$ and $p_{z}^{2}.$ The term $p^{2}/3$ in Eq. (%
\ref{7}) ensures that ${I}[p_{x}^{2}+p_{y}^{2}+p_{z}^{2}]=0$.

Since the scattering rate depends on the local number density \cite{henrik
book}, the mode damping involves a spatial average of the relaxation time.
(The simplest functional dependence one can assume for the collision rate is
that of the equilibrium distribution function.) To obtain collective modes,
we insert the trial function (\ref{5}) in the Boltzmann equation (\ref{8})
and then take moments of this equation by multiplying it by $%
f^{(0)}(1+f^{(0)})$ and by each of the terms in $\psi $ and then integrating
over $\mathbf{p}$ and $\mathbf{r}$ taking into account Eq. (\ref{7}). For
each moment we get a separate equation and thus obtain nine coupled
equations. The condition for the existence of nontrivial solutions results
in the following dispersion relation for the frequency of the collective
modes: 
\begin{equation}
\begin{array}{l}
\begin{array}{c}
\left( \omega ^{6}-4\omega _{a}^{2}\omega ^{4}+16\omega _{b}^{4}\omega
^{2}-64\omega _{c}^{6}\right) \\ 
+\frac{2}{3}i\left( \omega ^{6}-10\omega _{a}^{2}\omega ^{4}+32\omega
_{b}^{4}\omega ^{2}-96\omega _{c}^{6}\right) /(\omega \tau )
\end{array}
\\ 
-\frac{1}{3}\left( \omega ^{6}-8\omega _{a}^{2}\omega ^{4}+20\omega
_{b}^{4}\omega ^{2}-48\omega _{c}^{6}\right) /(\omega \tau )^{2}=0,
\end{array}
\label{12}
\end{equation}
where 
\[
\omega _{a}^{2}=\omega _{1}^{2}+\omega _{2}^{2}+\omega _{3}^{2}, 
\]
\[
\omega _{b}^{4}=\omega _{1}^{2}\omega _{2}^{2}+\omega _{1}^{2}\omega
_{3}^{2}+\omega _{2}^{2}\omega _{3}^{2}, 
\]
and 
\[
\omega _{c}^{6}=(\omega _{1}\omega _{2}\omega _{3})^{2}, 
\]
with 
\begin{equation}
{\frac{1}{\tau }}=\frac{\int d{\mathbf{r}}\tau ^{-1}({\mathbf{r}})\int
d{\mathbf{p%
}}f^{(0)}(1+f^{(0)})\varepsilon ^{2}}{\int d{\mathbf{r}}\int d{\mathbf{p}}%
f^{(0)}(1+f^{(0)})\varepsilon ^{2}},  \label{11}
\end{equation}
where $\varepsilon =p^{2}/2m$ is the kinetic energy. Equation (\ref{12}) is
our main result, which describes the frequency and attenuation of collective
modes in the collisionless and hydrodynamic regimes, as well as in the
intermediate regime.

\section{Mode Frequencies and Damping Rates for a trap with an axis of
symmetry}

\label{axial} In traps with an axis of symmetry the oscillations in the
axial and radial directions are decoupled and therefore the projection of
angular momentum $m$ about the symmetry axis of the trap is a good quantum
number and can be used to label modes. In this case the modes described for
a general trap in subsection \ref{sub1} have the following three forms: $i)$
The oscillations in the radial and axial directions are in phase (In the
spherically symmetric case this corresponds to the breathing mode, $a=b=c$). 
$ii)$ The oscillations in the two radial directions are out-of-phase with
those in the axial direction ( $a=b=-c/2$ for the case of spherical
symmetry). $iii)$ The oscillations in the two perpendicular radial
directions are out-of-phase with respect to each other and there is no
oscillation in the axial direction ($a=-b,$ $c=0$). Both of the first and
second modes have $m=0$, while for the last one $m=2$. For the rest of this
paper we denote the first mode as the $0^{+}$-mode, the second one as the $%
0^{-}$-mode, and the third one as the $2$-mode.

\subsection{Frequencies and Damping Rates}

The roots of Eq. (\ref{12}) can be obtained analytically as functions of $%
\tau $ and the characteristic frequencies $\omega _{a}$, $\omega _{b}$, and $%
\omega _{c}$, but they have a complicated functional dependence on the
parameters. However, for experimental traps with axial symmetry
characterized by $\omega _{1}$ $=\omega _{2}=\omega _{0}$ and $\omega
_{3}=\lambda \omega _{0},$ where $\lambda $ is the anisotropy ratio, the
roots simplify considerably. In this case the dispersion relation (\ref{12})
takes the form 
\begin{equation}
\omega \left[ \left( \omega ^{2}-2\omega _{0}^{2}\right) -i\omega \tau
\left( \omega ^{2}-4\omega _{0}^{2}\right) \right] \left[ \left( \omega
^{2}-r_{-}\omega _{0}^{2}\right) \left( \omega ^{2}-r_{+}\omega
_{0}^{2}\right) -i\omega \tau \left( \omega ^{2}-4\omega _{0}^{2}\right)
\left( \omega ^{2}-4\lambda ^{2}\omega _{0}^{2}\right) \right] =0,
\label{disp axial}
\end{equation}
where 
\begin{equation}
r_{\pm }=\frac{5+4\lambda ^{2}}{3}\pm \frac{1}{3}\sqrt{25-32\lambda
^{2}+16\lambda ^{4}}.  \label{rpm}
\end{equation}
Note that the dispersion relation in this case factorizes into two terms
(apart from the $\omega $ factor), one of which is independent of $\lambda $%
. This factorization can be understood by finding the eigenmodes that
correspond to the roots of (\ref{12}). As we shall discuss in more detail
below, the $\lambda $-independent term corresponds to the $2$-mode, in which
there is no oscillation in the $z$-direction. The other two modes,
corresponding to the $\lambda $-dependent factor, are the $0^{+}$- and $%
0^{-} $-modes. There are three modes of frequency $\omega =0$ in the
collisionless limit which correspond to thermal expansion. The first
corresponds to constant temperature change in all directions. This mode is
associated with the over-all factor $\omega $ in Eq. (\ref{disp axial}). The
second mode has $m=0$ and frequency $\omega =-i/2\tau $ in the collisionless
limit ($\omega _{0}\tau \gg 1$), and $\omega =-i/\tau $ in the hydrodynamic
limit ($\omega _{0}\tau \ll 1$). The third mode has $m=2$ and the same
values of frequency in the collisionless and hydrodynamic limits as for the
previous mode.

We start the analysis of (\ref{disp axial}) by looking at mode frequencies
in the collisionless and hydrodynamic limits. In the collisionless limit, $%
\omega \tau \gg 1$, we get: 
\begin{equation}
\omega \equiv \omega _{C}=2\omega _{0},2\omega _{0},2\lambda \omega _{0},\ \ 
\label{w2}
\end{equation}
while in the hydrodynamic limit, $\omega \tau \ll 1$: 
\begin{equation}
\omega \equiv \omega _{H}=\sqrt{2}\omega _{0},\;\sqrt{r_{-}}\omega _{0},\;%
\sqrt{r_{+}}\omega _{0},  \label{w hydro}
\end{equation}
where $\omega _{C}$ denotes the frequency in the collisionless limit and $%
\omega _{H}$ is the corresponding frequency in the hydrodynamic limit. Thus
we obtain the simple picture that the three collisionless modes of
frequencies $2\omega _{0},$ $2\omega _{0}$ and $2\lambda \omega _{0}$
correspond to the three hydrodynamic modes of frequencies $\sqrt{2}\omega
_{0},\sqrt{r_{-}}\omega _{0}$ and $\sqrt{r_{+}}\omega _{0},$ respectively.
(This is true only for $\lambda >1$. When $\lambda <1$ the mode of frequency 
$2\lambda \omega _{0}$ corresponds to the mode of frequency $\sqrt{r_{-}}%
\omega _{0}$ instead of that of frequency $\sqrt{r_{+}}\omega _{0}$.). We
note here that our result (\ref{w hydro}) agrees with previous works based
on either the conservation laws \cite{notes}, or taking moments of the
kinetic equation \cite{5string}.

To obtain frequencies of modes and their attenuation in the intermediate
regime, we insert\ $\omega =\omega _{r}-i\omega _{i}$ in Eq. (\ref{disp
axial}), and thus obtain two coupled equations for the real and imaginary
parts. 
In Figs. 1 and 2 we plot the frequency and the damping rate of the $0^{-}$%
-mode for values of $\lambda $ that satisfy the experimentally relevant
condition $\lambda \ll 1$.

It is instructive, also, to investigate the functional dependence of the
damping rate on $\lambda $ and $\tau $ in the collisionless and hydrodynamic
limits. This behavior can be deduced using the property that in both of
these limits the imaginary part $\omega _{i}$ is small, and thus one can
expand Eq. (\ref{disp axial}) in $\omega _{i}$. Let us first consider the
dispersion relation corresponding to the first factor in this equation. In
the collisionless limit, $\omega \tau \gg 1$, we can expand this factor to
first order in $1/(\omega \tau ),$ and write the solution as 
\begin{equation}
\omega =2\omega _{0}-i\Gamma _{2}^{c},  \label{x2}
\end{equation}
where $\Gamma _{2}^{c}$ is the small damping rate associated with the $2$%
-mode$.$ To first order in $1/\tau $ we obtain 
\begin{equation}
\Gamma _{2}^{c}=\frac{1}{4}\tau ^{-1}.  \label{x3}
\end{equation}
Similarly, for the other two modes associated with the $\lambda $-dependent
factor of (\ref{disp axial}) we obtain 
\begin{equation}
\Gamma _{0^{+}}^{c}=\frac{1}{12}\tau ^{-1},  \label{d1}
\end{equation}
and 
\begin{equation}
\Gamma _{0^{-}}^{c}=\frac{1}{6}\tau ^{-1}.
\end{equation}
In the hydrodynamic limit, $\omega \tau \ll 1,$ we expand in $\omega \tau $
and then insert $\omega =\sqrt{2}\omega _{0}-i\Gamma _{2}^{h}$ to get 
\begin{equation}
\Gamma _{2}^{h}=\omega _{0}^{2}\tau ,  \label{x4}
\end{equation}
together with 
\begin{equation}
\Gamma _{0^{+}}^{h}=\frac{(r_{+}-4\lambda ^{2})(r_{-}-4)}{2(r_{-}-r_{+})}%
\omega _{0}^{2}\tau ,  \label{xx}
\end{equation}
and 
\begin{equation}
\Gamma _{0^{-}}^{h}=\frac{(r_{-}-4\lambda ^{2})(r_{-}-4)}{2(r_{+}-r_{-})}%
\omega _{0}^{2}\tau .  \label{yy}
\end{equation}
We note that the damping rates in the collisionless limit are proportional
to $1/\tau $ and independent of $\lambda $.

Finally, we end this section by discussing some interesting features of the
hydrodynamic damping rates $\Gamma _{0^{+}}^{h}$ and $\Gamma _{0^{-}}^{h}$
given by Eqs. (\ref{xx}) and (\ref{yy}). This can be done by plotting these
rates as a function of $\lambda $ as shown in Fig. 3. From this figure, we
first note that $\Gamma _{0^{+}}^{h}$ vanishes for $\lambda =1.$ This is in
agreement with the well established fact that in an isotropic harmonic
potential the breathing mode does not relax, as was first shown by Boltzmann 
\cite{bol}. We also note that $\Gamma _{0^{-}}^{h}$ has the maximum value $%
1.13\omega _{0}^{2}\tau $ for $\lambda \approx 1.20.$ In the limit $\lambda
\longrightarrow 0$ (cigar-shaped cloud; a quasi one-dimensional system) the
damping rate vanishes, while in the limit $\lambda \longrightarrow \infty $
(disk-shaped cloud; a quasi two-dimensional system) the damping rate
approaches a constant value equal to $(3/4)\omega _{0}^{2}\tau $.

\section{Discussion}

\label{disc}
The damping rates found above were all expressed in terms of
the phenomenological time $\tau $, Eq. (\ref{11}). We shall now identify the
damping rates found above with the precise expressions obtained by a
microscopic calculation. If we assume that the collision rate is
proportional to the local density and take the classical limit, $f^{(0)}\ll
1 $, the spatial average entering Eq. (\ref{11}) may readily be performed,
resulting in $\tau =2\sqrt{2}\tau (0),$ where $\tau (0)$ is the relaxation
time at the center of the trap$.$
$\;$The damping rate,
$\Gamma _{0^{+}},$ for the $0^{+}$-mode$,$ which is given
by Eq. (\ref{d1}), therefore becomes $1/(24\sqrt{2}\tau (0))$. In Ref. \cite
{george2} this damping rate was calculated from a variational solution to
the Boltzmann equation in the collisionless limit, with the result 
\begin{equation}
\Gamma _{0^{+}}^{c}\leq \frac{1}{24\sqrt{2}}\frac{1}{\tau _{var}(0)}\left( 1+%
\frac{3}{16}\zeta (3)\left( \frac{T_{BEC}}{T}\right) ^{3}\right) ,
\label{rr}
\end{equation}
where the quantity in the parenthesis results from including degeneracy
effects to lowest non-vanishing order. Here $\tau _{var}(0)$ is the
variational viscous relaxation time given by $\tau _{var}^{-1}(0)=(4\sqrt{2}%
/5)n(0)\sigma \overline{v}$ \cite{henrik book}, where $n(0)$ is the central
density, $\sigma =8\pi a^{2}$ is the total cross section, $\overline{v}=%
\sqrt{8k_{B}T/(\pi m)}$ is the mean thermal velocity, and $T_{BEC}$ is the
Bose-Einstein transition temperature. The use of an improved trial function
changes the number $1/(24\sqrt{2}\tau (0))$ in Eq. (\ref{rr}) by only $\sim
1\%$ \cite{glast}$.$ A comparison between the damping rate Eq. (\ref{rr})
and Eq. (\ref{d1}) thus reveals that 
\begin{equation}
\tau (0)=\left( 1+\frac{3}{16}\zeta (3)\left( \frac{T_{BEC}}{T}\right)
^{3}\right) ^{-1}\tau _{var}(0).  \label{20}
\end{equation}
The same relationship, Eq. (\ref{20}), applies to the other two modes as
well. For the two experiments of Refs. \cite{{ket},{ket1}} our phenomenological
time, $%
\tau (0)$, in the center of the cloud is therefore given by $\tau (0)$ $%
\simeq 0.97\tau _{var}(0)$ for the experiment performed at $T\simeq 2T_{BEC}$%
, and $\tau (0)\simeq 0.82\tau _{var}(0)$ at $T\simeq T_{BEC}$.

In the hydrodynamic limit, an accurate microscopic calculation of the
collision rate is difficult to obtain due to the failure of the hydrodynamic
conditions near the surface of the cloud. This is because the mean free path
is too large for hydrodynamics to be valid. To solve this problem a cut-off
procedure was introduced in Ref. \cite{george1}, taking the hydrodynamic
description to be valid to a distance from the center of the cloud such that
an atom incident from outside the cloud has a probability less than $1/e$ of
not colliding with another atom. This cut-off procedure led to a logarithmic
dependence of the damping on the quantity $\alpha =\sigma n(0)\sqrt{%
k_{B}T/(2m\lambda \omega _{0}^{2})}$.
The leading term of this logarithmic dependence gives a damping rate
proportional to $[\ln {\alpha }]^{3/2}$ \cite{george1}.
The trial function (%
\ref{5}) is not capable of describing accurately the hydrodynamic limit,
since the very concept of hydrodynamic flow loses its meaning in the surface
region. Expressed in other terms, the deviation function has a very
different form in the outermost part of the cloud than it does in the
center, and this effect is not captured by our simple trial function.

The microscopic calculations performed in Refs. \cite{{george2},{george1}}
accounted for the damping rates only in the collisionless and hydrodynamic
limits. To compare with experiments, which are performed under intermediate
conditions, a simple phenomenological interpolation formula was suggested 
\cite{george2}. In the following we discuss the validity of this formula. In
particular we show that the interpolation formula is identical to our result
for the $2$-mode. Our results for the $0^{+}$- and $0^{-}$-modes cannot in
general be expressed in this simple form, but show a close resemblance, as
we shall demonstrate below. The interpolation formula suggested in Ref. \cite
{george2} is 
\begin{equation}
\omega ^{2}=\omega _{C}^{2}+\frac{\omega _{H}^{2}-\omega _{C}^{2}}{1-i\omega
\tau }.  \label{x13}
\end{equation}
By construction, the limits $\omega \tau \gg 1$ and $\omega \tau \ll 1$
yield the appropriate frequencies in the collisionless and hydrodynamic
limits, respectively. The $2$-mode is $\lambda $-independent, thus one can
easily put the corresponding dispersion relation (resulting from equating
the cubic factor in $\omega $ in Eq. (\ref{disp axial}) to zero) in the same
form as (\ref{x13}). For the $0^{-}$-mode we consider the two limiting cases 
$\lambda \ll 1$ and $\lambda \gg 1$, and the corresponding dispersion
relation can be written as 
\begin{equation}
\omega ^{2}=\omega _{C}^{2}+\frac{\omega _{H}^{2}-\omega _{C}^{2}}{1-i\omega
(6\tau /5)},{\ \ \ \ \ }(m=0,\ \lambda \ll 1),  \label{int41}
\end{equation}
\begin{equation}
\omega ^{2}=\omega _{C}^{2}+\frac{\omega _{H}^{2}-\omega _{C}^{2}}{1-i\omega
(4\tau /3)},{\ \ \ \ \ }(m=0,\ \lambda \gg 1).  \label{30}
\end{equation}
The formula corresponding to the $0^{+}$-mode in these limits takes the same
form as (\ref{x13}), but differs slightly from it, when $\lambda $ is
comparable to 1. To illustrate how well the simple interpolation formula,
Eq. (\ref{x13}), agrees with our results obtained from the solution of Eq. (%
\ref{disp axial}) we compare in Fig. 4 our calculated damping rate of the $%
0^{-}$-mode, for the case $\lambda =\sqrt{8}$, with the interpolation
formula.

From experiments one obtains independent measurements of the mode frequency $%
\omega _{r}$ and its damping rate $\omega _{i}$. Since our calculated values
of these quantities are functions of the collision rate, we can eliminate $%
\tau $ and obtain a relation between the real and imaginary parts. This
allows us to calculate the damping rate of a mode without knowing about the
characteristics of the collision processes leading to damping. Substituting $%
\omega =\omega _{r}-i\omega _{i}$ in Eq. (\ref{int41}) and eliminating $\tau 
$ from the two resulting equations leads to

\begin{equation}
\omega _{i}^{2}=\frac{1}{2}(\omega _{C}^{2}-3\omega _{H}^{2})-\omega
_{r}^{2}+\frac{1}{2}\sqrt{\omega _{C}^{4}-10\omega _{C}^{2}\omega
_{H}^{2}+9\omega _{H}^{4}+16\omega _{H}^{2}\omega _{r}^{2}}.  \label{x14}
\end{equation}
It is interesting to note that this expression is also what one obtains from
the simple interpolation formula (\ref{x13}). In fact the expression Eq. (%
\ref{x14}) is valid for any number replacing the ratio $6/5$ in Eq. (\ref
{int41}). In Fig. 4 we use Eq. (\ref{x14}) to plot $\omega _{i}$ as a
function of \ $\omega _{r}.$ Two values of $\lambda $ are used, namely $%
0.076 $ and $0.074,$ corresponding to the two experiments of Refs.
\cite{{ket},{ket1}}.
It can be clearly seen in this figure that the conditions of these
experiments are intermediate between the collisionless and hydrodynamic
limits. The deviation of \ Eq. (\ref{x14}) from the exact solution obtained
from Eq. (\ref{disp axial}) is shown in Fig. 5 where we plot $\omega _{i}$
as a function of $\omega _{r}$ for $\lambda =\sqrt{8}.$

The present method of obtaining the frequency and attenuation of collective
modes can be extended to higher modes of \ frequency $l\omega _{i},$ where $%
l $ is a positive integer, by employing trial functions that differ from the
one we have used in Eq. (\ref{5}), for the $2\omega _{i}$-modes.

\section{Conclusion}

\label{conc} Starting from the semiclassical Boltzmann equation and using
the relaxation time approximation we have derived expressions for the
frequencies and damping rates of the low-lying oscillatory modes of a dilute
Bose gas above the Bose-Einstein transition temperature. These expressions
give frequencies and damping rates as functions of a phenomenological
collision time and hence describe modes in the regime intermediate between
the collisionless and hydrodynamic regimes. Our results justify the use of a
simple interpolation formula for connecting the hydrodynamic with the
collisionless regime. The approach used in this work may be readily
generalized to higher-frequency modes. After the completion of the present
work we received a preprint by Gu\'{e}ry-Odelin et al. \cite{last} who
consider the $0^{+}$- and $0^{-}$-modes in the classical limit and obtain
conclusions in agreement with ours, and also show that the results are in
accord with the results of numerical simulations.

The authors would like to acknowledge G. Kavoulakis and W. Ketterle  for
helpful discussions.


\newpage
{\large Captions of Figures}
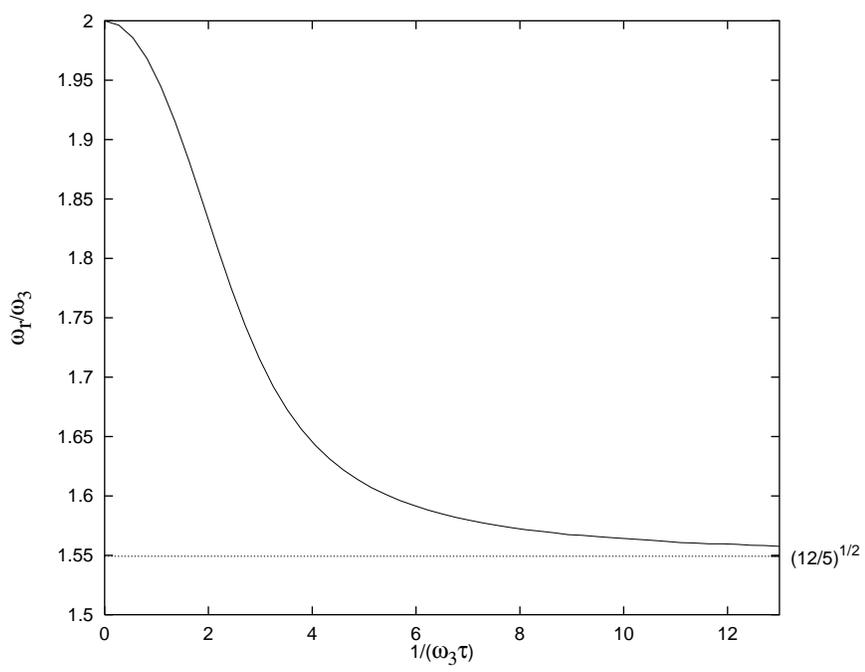
\begin{figure}[p]
\begin{center}
\end{center}
\caption{Frequency of the $0^{-}$-mode. This plot is obtained by taking the
real part of the lower solution ($\protect\sqrt{r_{-}}\protect\omega_{0}$)
of the dispersion relation resulting from equating the $\protect\lambda $%
-dependent factor in Eq. (\ref{disp axial}) to zero in the limit $\protect%
\lambda\ll 1$. The imaginary part is shown in Fig. 2.}
\end{figure}
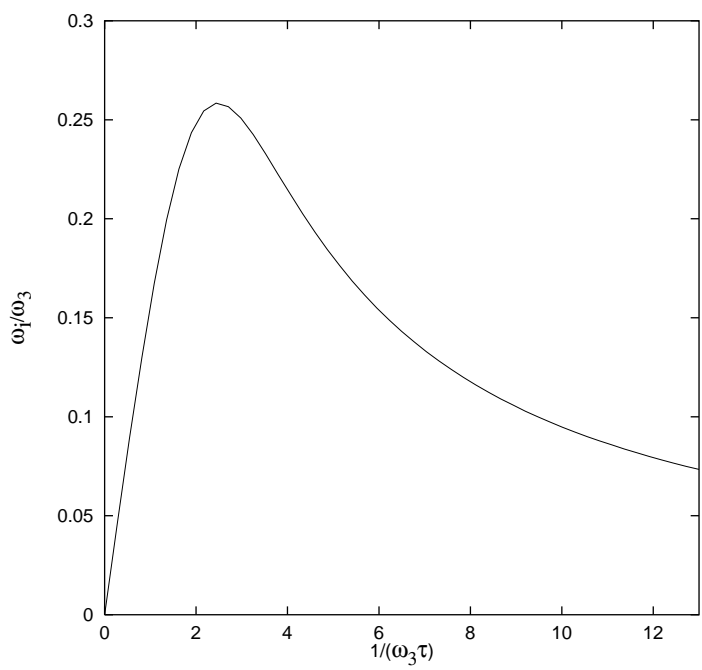
\begin{figure}[p]
\begin{center}
\end{center}
\caption{Damping rate of the $0^{-}$-mode, in the limit $\protect\lambda \ll
1.$\hspace{10cm}}
\end{figure}
\begin{figure}[p]
\begin{center}
\end{center}
\caption{The hydrodynamic damping rates $\Gamma _{0^{-}}^{h}$ and $\Gamma
_{0^{+}}^{h}$ in units of $\protect\omega _{0}^{2}\protect\tau ,$ as a
function of $\protect\lambda $. The rate $\Gamma _{0^{-}}^{h}$ is associated
with the $0^{+}$-mode while $\Gamma _{0^{+}}^{h}$ corresponds to the $0^{+}$%
-mode.}
\end{figure}
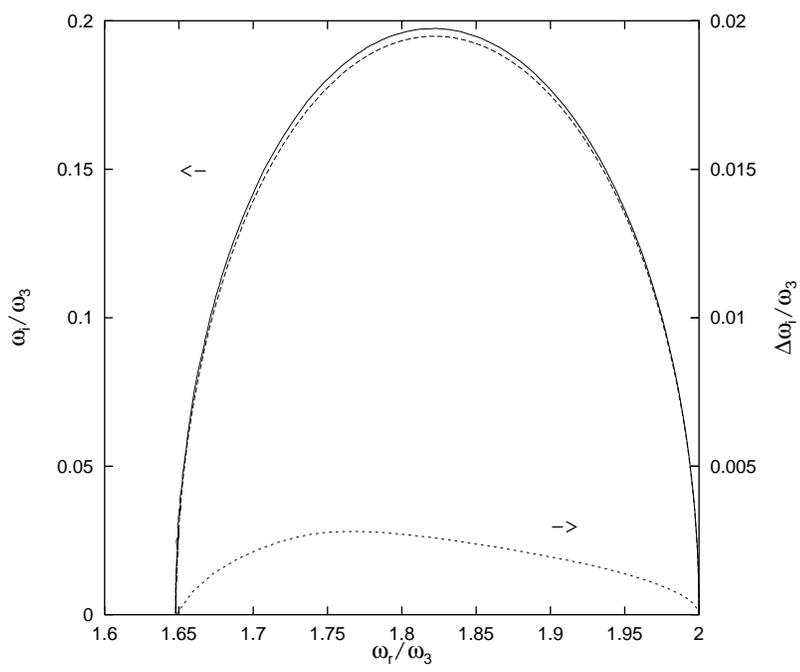
\begin{figure}[p]
\begin{center}
\end{center}
\caption{Damping rate versus mode frequency of the $0^{-}$-mode for $\protect%
\lambda =\protect\sqrt{8}$. The solid line represents the exact solution of
Eq. (\ref{disp axial}) and the dashed-dotted line is given by Eq. (\ref{x14}%
). The dotted line is the difference between the two upper curves (note the
change of scale).}
\end{figure}
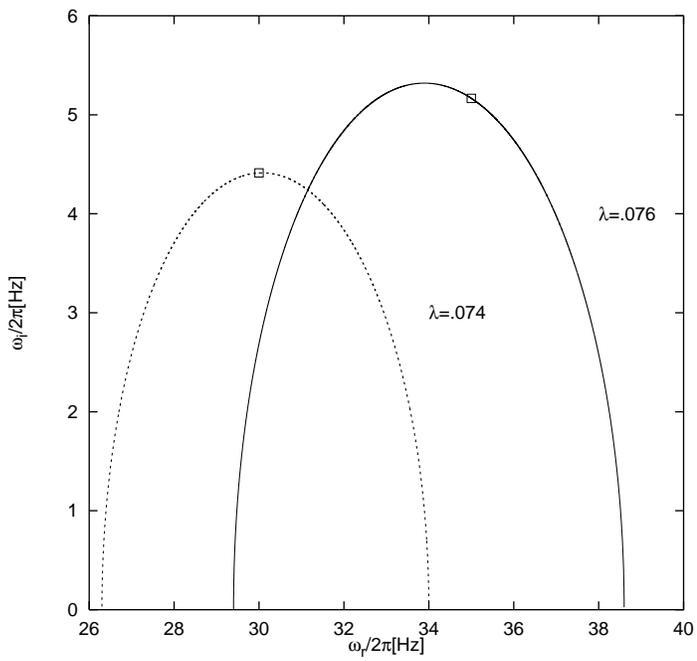
\begin{figure}[p]
\begin{center}
\end{center}
\caption{Damping rate versus mode frequency of the $0^{-}$-mode. The two
curves are given by Eq. (\ref{x14}) for different parameters corresponding
to the two experiments of Refs. [8,9]. The points indicate experimental
results.}
\end{figure}


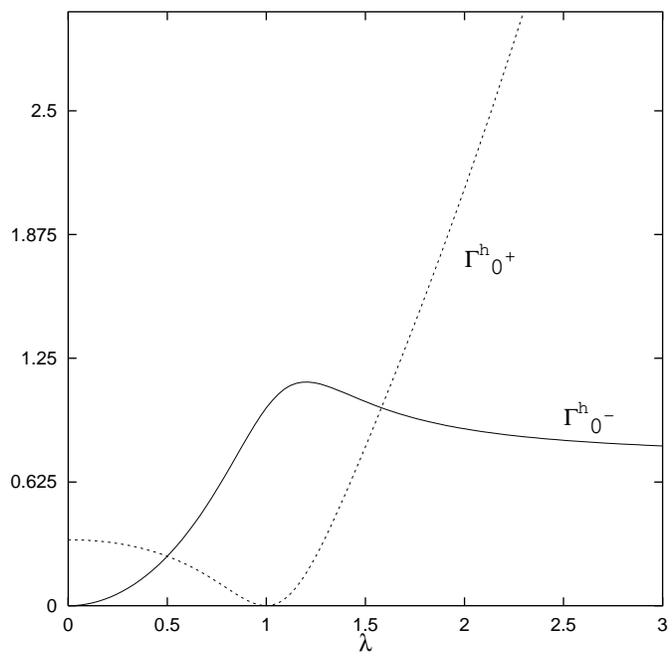
\end{document}